\documentclass[conference]{IEEEtran}
\usepackage[noadjust]{cite}
\usepackage{booktabs}
\usepackage{multicol,amsmath,amssymb,amsfonts}
\usepackage{algorithmic}
\usepackage{graphicx}
\usepackage{textcomp}
\usepackage{xcolor}
\def\BibTeX{{\rm B\kern-.05em{\sc i\kern-.025em b}\kern-.08em
    T\kern-.1667em\lower.7ex\hbox{E}\kern-.125emX}}
\begin{document}

\title{An Innovative Tool for Uploading/Scraping Large Image Datasets on Social Networks\\
}

\author
{
\IEEEauthorblockN{Nicolò Fabio Arceri}
\IEEEauthorblockA{\textit{Dept. of Math and Computer Science} \\
\textit{University of Catania}\\
Catania, Italy \\
50000-0001-9475-2382}
\and
\IEEEauthorblockN{Oliver Giudice}
\IEEEauthorblockA{\textit{Dept. of Math and Computer Science} \\
\textit{University of Catania}\\
Catania, Italy \\
0000-0002-8343-2049}
\and
\IEEEauthorblockN{Sebastiano Battiato}
\IEEEauthorblockA{\textit{Dept. of Math and Computer Science} \\
\textit{University of Catania}\\
Catania, Italy \\
0000-0001-6127-2470}
}

\maketitle

\begin{abstract}
Nowadays, people can retrieve and share digital information in an increasingly easy and fast fashion through the well-known digital platforms, including sensitive data, inappropriate or illegal content, and, in general, information that might serve as probative evidence in court. Consequently, to assess forensics issues, we need to figure out how to trace back to the posting chain of a digital evidence (e.g., a picture, an audio) throughout the involved platforms—this is what Digital (also Forensics) Ballistics basically deals with. With the entry of Machine Learning as a tool of the trade in many research areas, the need for vast amounts of data has been dramatically increasing over the last few years. However, collecting or simply find the “right” datasets that properly enables data-driven research studies can turn out to be not trivial in some cases, if not extremely challenging, especially when it comes with highly specialized tasks, such as creating datasets analyzed to detect the source media platform of a given digital media. In this paper we propose an automated approach by means of a digital tool that we created on purpose. The tool is capable of automatically uploading an entire image dataset to the desired digital platform and then downloading all the uploaded pictures, thus shortening the overall time required to output the final dataset to be analyzed.
\end{abstract}


\section{Introduction}

With the entry of Machine Learning as a tool of the trade in many research areas, the need for vast amounts of data has been dramatically increasing over the last few years. However, collecting or simply find the “right” datasets that properly enables data-driven research studies can turn out to be not trivial in some cases, if not extremely challenging, especially when it comes with highly specialized tasks, which often mandatorily lead to the creation of ad hoc datasets.

One of the research fields where this problem happens to be particularly serious is Digital Forensics, which has led to the creation of public datasets like SHADE \cite{b1}, VISION \cite{b2}, RAISE \cite{b3}, and R-SMUD \cite{b4}, all of which proposed to deal with many typical tasks of the discipline. We will be discussing more about Digital Forensics and related tasks in the next sub-section (Digital Forensics Background).

One of the tasks where collecting large datasets can happen to be utterly prohibitive is the source media platform detection problem (also known as source social network identification task \cite{b5}), that is detecting the platform of provenance (e.g., Facebook, Skype, Mastodon) of a given digital content, such as a picture, a video, or an audio content. To better understand how long it could be taking collecting data for this kind of task, just picture how intensive it could get to be the manual process of uploading and downloading thousands and thousands of image files, one by one (or so), to and from social media or even instant messaging platforms, simply to obtain a dataset that is large enough to get significant results with a Machine Learning approach. Clearly, an automated procedure rather than a manual one would be ideal for the purpose, and that is exactly what we propose in this paper. Before introducing further details about our proposal, it is worth laying out some background about Digital Forensics and the main problem addressed by this discipline.

\subsection{Digital Forensics Background}

Forensic Science (sometimes shortened to Forensics) is \cite{b6} the application of technical and scientific methods to the justice, investigation and evidence discovery domain. Many fields of science are finding new applications in Forensic Science, giving investigators more powerful tools.

One of the Forensics fields with many successful results is Digital Forensics: Computer Science meets Forensics. Nowadays, Digital Forensics has developed into new specific fields: Computer Forensics, Disk Forensics, Network Forensics, Image Forensics, etc. More specifically, by using Image Processing science and domain expertise, Image Forensics analyses an image to detect forgeries\footnote{A forgery can be any modifications that occurred to a source image, without any special negative or positive meaning.} or manipulations (Image Source Forensics, i.e., image integrity/authenticity verification) and to reconstruct the history of an image since its acquisition (Image Ballistics) \cite{b6}.

When it comes to social media platforms, Image Source Forensics and Image Ballistics also come on stage. Indeed, recent studies have shown that such platforms alter images for bandwidth, storage, and layout reasons \cite{b7} as a result of the uploading action. Besides that, studies in \cite{b8}, \cite{b9} also show that this alterations process leaves a sort of fingerprint on the JPEG image format. This evidence can be exploited to understand if the image was uploaded to a particular social media platform, a task also known as the “source social media platform identification”. This very task can be used to extend the source identification test to more than one platform at the same time over the same picture; and that is where Image Ballistics comes in. In other words, Image Ballistics exploits social-media-left fingerprints to trace back to the posting/sharing chain of a picture throughout the platforms involved. In conclusion, when it comes to Social Media Platforms, Image Source Forensics and Image Ballistics combine to answer the following question: which platforms does that picture come from? Answering that question is important to address security and privacy issues, since these platforms contain sensitive and personal data of hundreds of millions of people and are also integrated into millions of other websites. For example, in \cite{b10} the authors investigate the defamation issue.

In the last few years, many researchers have applied data-driven approaches for forensics \cite{b11}, \cite{b12} inspired by the excellent performance obtained by deep learning and convolutional neural network approaches \cite{b13}–\cite{b15}. More specifically in the image forensics field, in \cite{b5} the authors present a survey about data-driven algorithms dealing with the problem of Image Source Forensics by dividing the area into five sub-topics: Source Camera Identification, Recaptured Image Forensic, Computer Graphics (CG) Image Forensic, GAN-Generated Image Detection, and, precisely, Source Social Media Platform Identification. For what concerns the latter, many important successful studies such as \cite{b8}, \cite{b16}–\cite{b22} were conducted in the last few years in order to produce not only detection criteria but real working detection engines.

\subsection{The Motivation about this Study}

To sum it up, when it comes to the source social media platform identification task, creating datasets to be analyzed is a long-lasting process due to the great deal of manual uploads and downloads required. So, in conclusion, the motivation for this study is all about speeding up that manual process.

To be precise, these would be the steps with the respective latency times to upload just a single picture to a single platform:

\vspace{+\topsep}
\begin{itemize}
\item time needed to login to the platform;
\item time needed to manually select the image (or a set of) to be uploaded for pre-loading purposes – this is an explicit requirement of any platform GUI;
\item time that elapses between an upload and another one of two images or two sets of images;
\item time that elapses between the upload and the download phase, the latter being particularly onerous due to the need to search and select the images to be downloaded via the platform GUI;
\item time necessary for the preparation of the directory structure inside the local machine intended to host the downloaded files;
\item time necessary for sorting the downloads into the appropriate and already prepared directories, which often involves manual interaction with the operating system GUI in order to select the correct sub-folder in which to pour the incoming files;
\item and, of course, we should take into account the hours of stop necessary for human being’s rest between one job session and the next.
\end{itemize}
\vspace{+\topsep}

And this is the work amount required for one platform only, which must be repeated for each platform we want to add. Clearly, this approach is not feasible.

In this paper we propose an automated approach by means of a digital tool that we created on purpose. The tool is capable of automatically uploading an entire image dataset to the desired digital platform and then downloading all the uploaded pictures, thus shortening the overall time required to output the final dataset to be analyzed. Despite the presence of other similar projects and publications \cite{b23}–\cite{b29} addressing the creation of tools to ease data analysis tasks, our tool aims to be rather unique, so ultimately innovative. Indeed, all of them are focused on extracting information or images, some by using tools like Instalooter \cite{b28} and BeautifulSoup, LXml, and RegEx \cite{b29}. However, none of them claim the ability to upload an initial (and so unaltered) dataset as our tool is capable of. Additionally, none of them claim their tools as being specially designed in order to be easy to extend toward further platforms. As will be shown in the next section, our tool is.

\section{Method}

The previous section basically states that creating a picture dataset for ballistic studies about reconstructing on-line media platform posting/sharing chains is a considerably time-consuming task. So, in order to address this problem, we decided to create a tool for automating the problem’s key points previously outlined, thus speeding up the whole process. Throughout this section we will then discuss the main tool’s features that make it possible to (1) effectively speed up the aforementioned process and (2) profitably support future digital ballistics related studies. We will also be describing a test report.

\subsection{Overview about the Tool’s Usage and Goals}

The tool we developed is a CLI (Command Line Interpreter) Python application. The basic idea behind this tool is as simple as visually showed in the Fig. 1, that is:

\vspace{+\topsep}
\begin{enumerate}
\item preparing the initial picture dataset to be uploaded by putting it in a dedicated special directory (more about that in a bit);
\item setting up a JSON config file describing where the dataset is located and which platforms the tool will have to be uploading the dataset to (plus some other options that we are caring to get into later);
\item launching the tool and waiting for it to return the final
dataset by downloading it after having uploaded to the
platforms specified inside the JSON config file.
\end{enumerate}
\vspace{+\topsep}


\begin{figure}[htbp]
\centerline{\includegraphics[width=240.12pt,height=141.84pt]{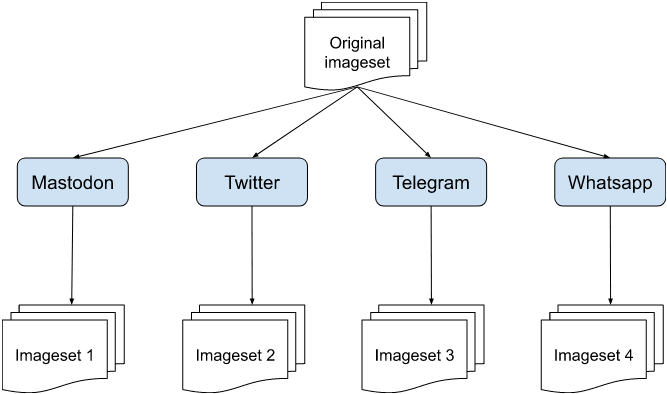}}
\caption{\textbf{THE EASIEST USE CASE} — At its most basic usage, the tool allows to send an initial image dataset to one or more platforms. After sending the dataset to them, it starts to download each of the uploaded pictures directly into a dedicated special directory structure inside a preset base folder.}
\label{fig1}
\end{figure}


The tool currently supports the following platforms: Skype, Reddit, Mastodon, Viber, Signal, Diaspora, Odnoklassniki (ok.ru), while Friendica, Imgur, Flickr, Twitter, Facebook post, Facebook messenger (so 13 platforms in overall) will be integrated in the very near future (we have already tested the respective APIs and ascertained their integration feasibility).

So, to sum it up, in this paper we propose a tool that is capable to upload and “scrape” (i.e., download) large image datasets in an automatic fashion to and from as many platforms as needed, so as to shorten the time required for the user to create the final dataset to be analyzed later. As already said, the tool takes as input a JSON configuration file through which the user can specify the (local) path of the source dataset and the platforms which work with.

\subsection{The pipelining feature}

The scenario in Fig.~\ref{fig1} is aimed to produce, for each of the listed platforms, an exclusive dataset preserving only the alterations coming from the respective platform. In other words, for instance, image dataset 1 (as shown in Fig.~\ref{fig1}) would contain the pictures altered solely by the platform named ”Mastodon”. But what if we wanted to include mixed alterations in a final dataset, that is alterations coming from more than one platform? So, here is where the pipelining feature comes.

Pipelining is the script’s ability of turning an output coming off a platform to an input for another platform, thus making the pictures of the final dataset preserve all the alterations applied all along the platforms connected that way. Fig.~\ref{fig2} shows three different examples of jobs in which one or more pipeline were included.

\subsection{The JSON config file}

The JSON configuration file provides a bunch of options that can be used to drive and customize the tool’s behavior. As an example, Fig.~\ref{fig3} shows a sample file implementing the
scheme reported in Fig.~\ref{fig2}b. However, discussing every single property available is out of the scope of this paper; instead, we will provide a summarization of what can be achieved by properly tinkering with the options available. Simply put, by tweaking over the config file settings, the following features can be handled/activated:

\vspace{+\topsep}
\begin{itemize}
\item setting up the dataset source folder;
\item defining one or more pipelining point;
\item setting up the sending pool size (see Fig.~\ref{fig3} for more details);
\item setting up the default action (amongst ask user, skip task, automatically fix the issue, and terminate the whole process) to handle critical issues that can occur during the process, such as network failures and rate limit reaching;
\item activating the debugging mode;
\item activating the multi-picture sending if available.
\end{itemize}
\vspace{+\topsep}

\subsection{Modular Architecture}

Automation clearly offers a faster-than-manual way to performing the job. But being able to integrate a new platform just as quickly as possible is valuable too. With that in mind we structured the tool’s code so that it would take not too long to integrate a new platform needed for later analysis purposes (see Fig.~\ref{fig4} for further details).

\subsection{Anatomy of the Output(s)}

The tool outputs basically three pieces of information: (1) the pictures processed (i.e., the final dataset), (2) a job map file, and (3) a job status file.

As seen in Fig.~\ref{fig3} while discussing the JSON config file, the whole process is organized into tasks, not simply platforms. Every task produces a dataset on its own, and this is stored
in a dedicated directory having the same name as the task, as shown in Fig.~\ref{fig5}.

A job map file is a JSON file that is being continuously updated during the job. It aims to keep track of various useful metadata related or binarily held inside any picture processed
by the tasks performed during the job, such as:

\vspace{+\topsep}
\begin{itemize}
\item the path of the picture produced by the task;
\item the relative source picture (either original or pipelined)
of the produced-by-the-task picture;
\item a failure-during-update flag;
\item Exif data, if any;
\item a picture’s hash;
\item the quantization matrix;
\item other data of interest.
\end{itemize}
\vspace{+\topsep}

A job status file is a JSON file that is as continuously updated as the map file. It aims to keep track of the number of pictures actually processed (either successfully or not). The reason for it to exist is allowing interrupted processes to be resumed.

\subsection{Experiment Description}

The tool was tested with the 1000-image version of the RAISE public dataset \cite{b3} on the following 6 platforms: Skype, Reddit, Mastodon, Signal, Diaspora, Odnoklassniki (ok.ru). The configuration file we used to perform the test is the one


\begin{table*}[htbp]
\centering
\begin{tabular}{ccc} 
\includegraphics[height=175pt]{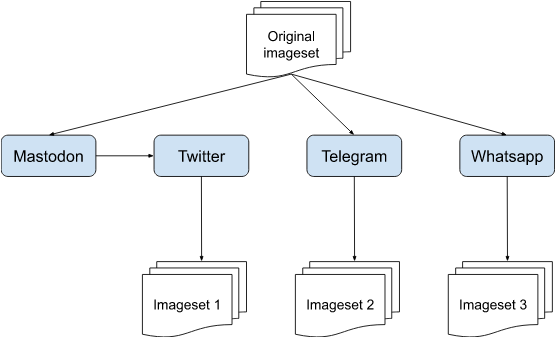} & \includegraphics[height=175pt]{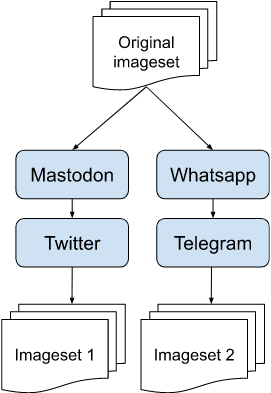} & \includegraphics[height=175pt]{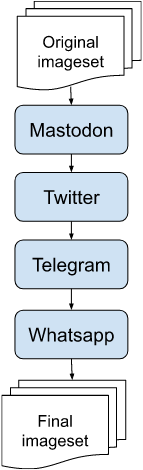} \\
(a) & (b) & (c) \\
\end{tabular}
\label{default}
\end{table*}
\begin{figure*}[htbp]
\caption{\textbf{PIPELINED USE CASES} — (a) Here we have just one pipeline defined, with ``Imageset 1'' being made up of pictures affected by alterations from both Mastodon and Twitter. On the other hand, ``Imageset 2'' and ``Imageset 3'' include alterations coming respectively from Telegram and Whatsapp. (b) Two independent pipelines are defined. (c) Only one pipeline is defined with four platforms involved.}
\label{fig2}
\end{figure*}


\begin{figure}[htbp]
\centerline{\includegraphics[width=240pt]{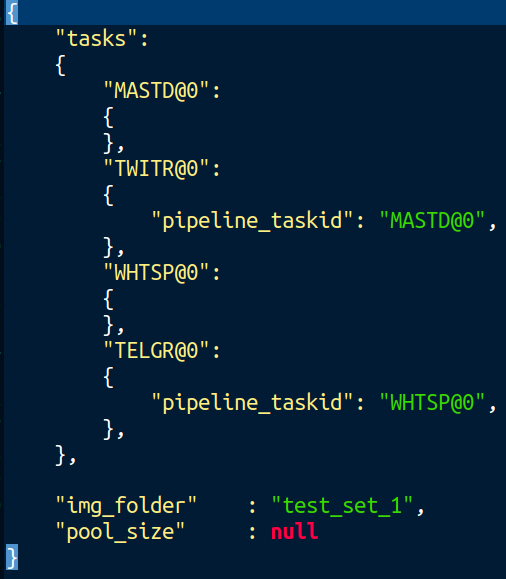}}
\caption{\textbf{THE JSON CONFIGURATION FILE} — Configuration file implementing the scheme reported in Fig.
2b. Each task is identified by an identifier in the form of PLATFORM NAMEPROGRESSIVE NUMBER. The property ``pipeline taskid'' is for declaring the incoming task, whereas the properties ``img folder'' is for specifying the source directory name (i.e., the directory which the source dataset resides in). A pool size (specified by the ``pool size'' property) is the number of pictures sent at once towards a single platform in case the multi-send functionality was activated for the running platform. When set to null, the tool uses the default pool size limit of the just running platform, which is defined inside its own worker class (see ”D. Modular Architecture” for more about this). If set, the tool will apply the same value for each platform instead.}
\label{fig3}
\end{figure}


shown in Fig.~\ref{fig6}, which is structured to serve a use case like the one described in Fig.~\ref{fig1}.

Since the initial dataset was made up of 1000 pictures, a total of 6000 pictures were expected (i.e., 1000 pictures by 6 platforms). The tool completed the job by uploading and downloading 6000 picture files for a total data volume of almost 17 GB (4 GB for the uploaded original dataset, 13 GB for the downloaded dataset). The process, apart from a handful of interruptions (due to the achievement of the rate limit for Mastodon) of the order of half an hour per interruption, lasted approximately 27 hours overall.

The test was carried out on a Lenovo Z50-70 / Lancer 5A5 laptop with an Intel ® Core TM i5-4210U CPU @ 1.70 GHz (1 physical processor, 2 core, 4 threads), 16.286.912 KiB ( 16 MiB) RAM, running Ubuntu 20.04.6 LTS as an operating system.

\section{Results and conclusions}

\subsection{Faster Picture Dataset Creation + Faster Platform Integration = Early Start of Dataset Analysis}

With a manual approach, assuming of being able to send 10 images at a time through the native loading GUI of the various platforms, we should be sending 1000/10 chunks for each platform session, i.e., 100 × 6 = 600 pictures sent in overall. If only 1 minute were taken for each of the 7 points of latency listed in regard of the manual approach, each chunk sending would take 7 minutes longer than the corresponding sending performed via the tool, generating a cascade delay of 600 × 7 = 4200 minutes (about 3 days) for the entire process, which is equal to 70 hours (about 3 days) more (manual) work to obtain the same dataset.


\begin{figure*}[htbp]
\centerline{\includegraphics[width=290pt]{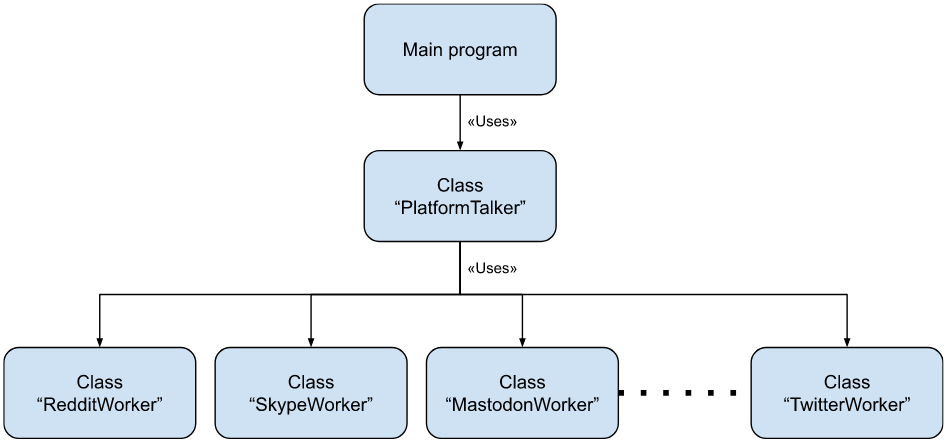}}
\caption{\textbf{TOOL’S ARCHITECTURE} - Worker classes are keys in achieving our fast integration goal. Each class requires four methods to be implemented, that are: connect(), upload proc(), download proc(), and disconnect(). Besides that, each method follows a quite stable design pattern, this allowing high quality in maintenance and extendibility as well as development speed according to well-known principles and best practices in Software Engineering.}
\label{fig4}
\end{figure*}


\begin{figure}[htbp]
\centerline{\includegraphics[width=190pt]{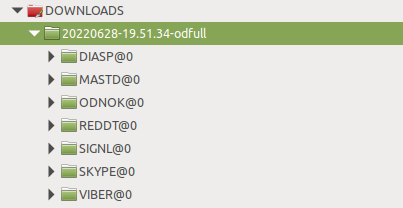}}
\caption{\textbf{JOB DIRECTORY STRUCTURE} — Every time a job is launched, a new root directory for the job is created to store the outputs related to it. Then, for each task defined inside the JSON config file, a dedicated directory is created inside the root directory of the job. At the end of the process, these directories will end up containing the pictures coming out of the respective tasks.}
\label{fig5}
\end{figure}


Furthermore, the following factors drastically worsen the manual approach in favor of the automated one. Here’s the reason why:

\vspace{+\topsep}
\begin{enumerate}
\item the limit of 10 images per GUI upload is rather optimistic;
\item the hypothesis of 1 minute per session is very optimistic, unlikely;
\item we haven’t taken into account the hours of break needed by human beings to rest between one session and the next one.
\end{enumerate}
\vspace{+\topsep}

This analysis, albeit qualitative, suggests that the first goal (i.e., effectively speeding up the process of creating a dataset of pictures coming from both social media and instant messaging platforms) is definitely reached and improvable as well.

Moreover, as previously discussed, the modular architecture of the tool ensures a low-effort extendibility and then effective rapid integration (three to seven days on average) of any APIs coming from platform still to be supported. So, not only execution speed is assured but also speed in enriching the tool capabilities, thus allowing an early start of any dataset analysis phases.

\subsection{Concretely Supporting Future Web-Posting/On-Line Sharing Ballistic Studies}

There are two important questions that need to be answered when it comes to posting/sharing ballistics:

\vspace{+\topsep}
\begin{enumerate}
\item is a certain digital media platform leaving some kind of fingerprint or marks/traces embedded within digital contents?
\item is it possible to reconstruct the chain of posting/sharing throughout certain digital media platforms?
\end{enumerate}


\begin{figure}[htbp]
\centerline{\includegraphics[width=130pt]{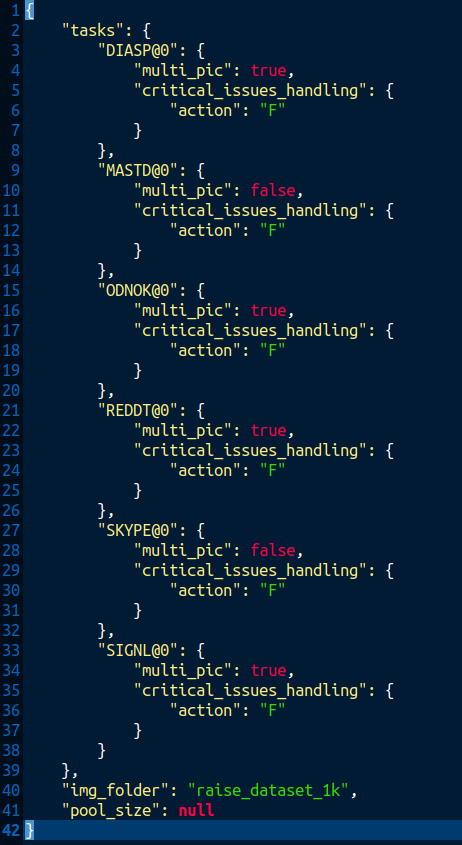}}
\caption{\textbf{THE CONFIGURATION FILE USED FOR THE EXPERIMENT} — At the first level, all the three
main properties are set, namely ”tasks”, ”img folder”, and ”pool size”. The first one contains all the sub-properties related to each actual task, namely ”MASTD@0” for Mastodon, ”ODNOK@0” for Odnoklassniki (ok.ru), ”REDDT@0” for Reddit, ”SKYPE@0” for Skype, and ”SIGNL@0” for Signal. When ”multi pic” is set to false, the tool will use the single picture sending mode. The ”critical issues handling” property
is intended to group properties allowing the user to select how to handle an error situation. As for now, the only way to manage a situation is by defining a default action. Specifically, an ”F” means ”try to automatically fix the issue; if it fails, ask the user about the action”.}
\label{fig6}
\end{figure}


In its most basic usage, our tool is trivially capable of building datasets that are useful to answer the first question. On the other hand, the pipelining feature allows to investigate about the second question. Hence even the second goal of this study (i.e., creating a tool that is useful for posting/sharing chain ballistics reasons) is indeed reached as well.

In light of what has been said, we can cite at least two research works that could have taken advantage of our tool, if it existed at their publication time:

\vspace{+\topsep}
\begin{enumerate}
\item \textbf{``A Classification Engine for Image Ballistics of Social Data''} \cite{b8}, where the authors used a custom dataset of pictures manually taken by using various cameras;
\item \textbf{``Multi-clue reconstruction of sharing chains for social media images''} \cite{b22}, where the authors used “R-SMUD” \cite{b4} as a source dataset for their experiment, that is a selection from RAISE \cite{b3}.
\end{enumerate}
\vspace{+\topsep}

\section{Future works}

We have identified two different directions for future works that could be enabled by our tool. The first one is about extending the tool itself. As previously said, its modular architecture will allow us to add more platforms in a relatively short time; that is, we count on hitting much more than just 13 platforms in the very near future. Of course, enriching the amount of information provided by the job map file is planned as well. The second direction is about exploiting our own tool in order to support our future data-driven-based studies, similarly to those conducted in \cite{b8}, \cite{b16}-\cite{b19}, \cite{b21}, \cite{b22}.

\end{document}